\documentclass[manuscript,screen]{acmart}
\AtBeginDocument{%
  \providecommand\BibTeX{{%
    \normalfont B\kern-0.5em{\scshape i\kern-0.25em b}\kern-0.8em\TeX}}}

\setcopyright{acmcopyright}
\copyrightyear{2018}
\acmYear{2018}
\acmDOI{XXXXXXX.XXXXXXX}

\acmConference[Conference acronym 'XX]{Make sure to enter the correct
  conference title from your rights confirmation emai}{June 03--05,
  2018}{Woodstock, NY}
\acmPrice{15.00}
\acmISBN{978-1-4503-XXXX-X/18/06}

\usepackage{titlesec}

\titleformat{\subsubsection}[hang]{\normalfont\normalsize\bfseries}
{\thesubsubsection}{0.3em}{}

\usepackage{tabularx}

\begin{document}

\title[Can a Funny Chatbot Make a Difference?]{Can a Funny Chatbot Make a Difference? Infusing Humor into Conversational Agent for Behavioral Intervention}

\author{XIN SUN\textsuperscript{1}, Isabelle Teljeur\textsuperscript{1}, Zhuying Li\textsuperscript{2}, AND Jos A. Bosch\textsuperscript{1} \\ 
{\small \textsuperscript{1}University of Amsterdam} \\ 
{\small \textsuperscript{2}Southeast University} 
}

\makeatletter
\def\ps@headings{%
    \def\@oddhead{}
    \def\@evenhead{}
    \def\@oddfoot{\normalfont\hfil\thepage\hfil}
    \def\@evenfoot{\normalfont\hfil\thepage\hfil}
}
\makeatother
\pagestyle{headings} 


\begin{abstract}
Regular physical activity is crucial for reducing the risk of non-communicable disease (NCD). With NCDs on the rise globally, there is an urgent need for effective health interventions, with chatbots emerging as a viable and cost-effective option because of limited healthcare accessibility. Although health professionals often utilize behavior change techniques (BCTs) to boost physical activity levels and enhance client engagement and motivation by affiliative humor, the efficacy of humor in chatbot-delivered interventions is not well-understood. This study conducted a randomized controlled trial to examine the impact of the generative humorous communication style in a 10-day chatbot-delivered intervention for physical activity. It further investigated if user engagement and motivation act as mediators between the communication style and changes in physical activity levels. 
66 participants engaged with the chatbots across three groups (humorous, non-humorous, and no-intervention) and responded to daily ecological momentary assessment questionnaires assessing engagement, motivation, and physical activity levels. 
Multilevel time series analyses revealed that an affiliative humorous communication style positively impacted physical activity levels over time, with user engagement acting as a mediator in this relationship, whereas motivation did not. 
These findings clarify the role of humorous communication style in chatbot-delivered physical activity interventions, offering valuable insights for future development of intelligent conversational agents incorporating humor.
\end{abstract}



\maketitle


\section{Introduction}

Noncommunicable diseases (NCDs) account for approximately 70\% of deaths worldwide, highlighting the critical need for effective interventions to mitigate these diseases. Physical activity stands out as a universally applicable and highly effective preventive approach. Increasing physical activity typically requires individuals to make lifestyle changes. To support these changes, health professionals often utilize Behavior Change Techniques (BCTs) during health counseling sessions \cite{michie2012identification,samdal2017effective}. 
These techniques encompass intervention-specific strategies, such as setting precise goals, celebrating small victories, or leveraging social support, each of which is grounded in psychological theories and supported by empirical evidences from behavioral science research. The application of these carefully crafted techniques enables health professionals to effectively promote and maintain higher levels of physical activity among individuals, contributing significantly to the global fight against NCDs \cite{dombrowski2012identifying, michie2009effective, michie2012theories}. 
While the BCTs employed in health interventions play a crucial role, a strong therapeutic alliance between health counselors and their clients is equally essential \cite{greenberg2003therapeutic, horvath1993role, ardito2011therapeutic}. Within this alliance, humor emerges as a significant tool, often integrated by healthcare professionals to enhance health message efficacy, foster engagement, and fortify the therapeutic relationship \cite{blanc2014humor}. This addition of humor into conversations lightens interactions, crafting a positive, participatory atmosphere for clients on their healthcare journey, subsequently improving their commitment to positive life changes, engagement, and satisfaction with the process \cite{sultanoff2013integrating,bennett2003humor,okun2014effective,shao2023empathetic}. 
Among various humor types, affiliative humor, characterized by jokes and lighthearted banter aimed at easing social interactions and building rapport, is universally appreciated due to its reliance on shared human experiences and emotions, making it accessible and relatable to diverse audiences. Its use in health counseling is particularly impactful, promoting comfortable, trusting, and engaging counselor-client relationships, thus enhancing client motivation \cite{dziegielewski2003humor,hussong2021use}. 

Recently, there has been an increasing interest in understanding how technology, especially the conversational agents or chatbots, can supplement the traditional healthcare providers in delivering interventions. Like humans, chatbots can effectively integrate BCTs into physical activity interventions, facilitating guidance through text-based interactions akin to the conversational support provided by human counselors \cite{wlasak2023supporting}. One of the key advantages of chatbots is their potential for wide scalability and cost-effectiveness, a significant consideration given the often prohibitive costs and limited availability of healthcare professionals. 
Research indicates that chatbot interventions employing BCTs have been successful in enhancing physical activity and minimizing sedentary behavior among users~\cite{kramer2019investigating,kramer2020components,sillice2018using,wlasak2023supporting}. Nonetheless, chatbots inherently lack the empathy, personal connection, and human touch characteristic of interactions with human counselors~\cite{kim2012anthropomorphism,pfeuffer2019anthropomorphic}. This limits the chatbots' potential to sustain user motivation, engagement, and emotional connection, which play a crucial role in mediating the relationship between the intervention and changes in physical activity~\cite{oosterveen2017systematic,teixeira2012exercise}.

\begin{figure}[htbp]
\centering
\includegraphics[width=0.75\textwidth]{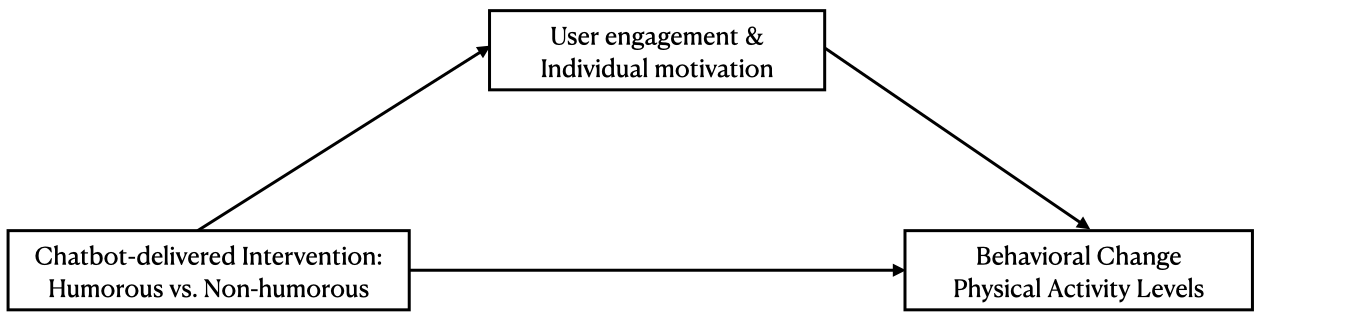}
\caption{Mediation Model. User engagement (H2a) and motivation (H2b) serve as mediators in the relationship between (humorous) communication style and physical activity levels (Hypothesis 2).}
\label{fig:mediator_model}
\end{figure}

To address these limitations, researchers are exploring innovative solutions. A study by Olafsson et al~\cite{olafsson2020motivating} reveals that infusing humor into the conversational agent for behavioral intervention can enhance user engagement and motivation. Participants engaged more with chatbots that employed humor, mirroring the increased engagement typically observed with human counselors. Therefore, integrating humor into chatbot interventions may compensate for the absence of empathy and personal connection, fostering a more engaging and motivating environment for users~\cite{okun2014effective,shao2023empathetic}. It is crucial to note that while humor itself is not a direct catalyst for change in physical activity, it significantly contributes to user engagement and motivation—key components indispensable for instigating actual behavioral change~\cite{glanz2015theory}.
However, existing research~\cite{olafsson2020motivating} primarily demonstrates that humorous chatbots can elevate user engagement and motivation without clarifying their real impact. The actual influence of a chatbot's communication style, especially its use of humor, on physical activity levels remains unassessed.
Technical constraints previously hindered the integration of appropriate and irrepetitive humor into chatbot dialogues, limiting the exploration of humor's long-term effects~\cite{olafsson2020motivating}. The advent of advanced large language models (LLMs), like ChatGPT~\cite{chatgpt}, eliminate these limitations. Such generative LLMs allow for the implementation of adaptive humorous conversations in chatbots and facilitates the examination of the long-term impact of humorous communication styles in chatbot-delivered behavioral interventions.
Hence, our research seeks to answer the following two research questions: 
\begin{itemize}
    \item How does a chatbot-delivered intervention with a humorous communication style influence users' physical activity levels~\textbf{(RQ1)}?
    \item Whether the user engagement and motivation mediate the relationship between a humorous communication style in chatbot-delivered interventions and changes in physical activity levels~\textbf{(RQ2)}?
\end{itemize}
We therefore hypothesized that the health behavioral intervention delivered by a chatbot with a humorous communication style will exert a stronger positive effect on physical activity levels than interventions using a non-humorous style (H1). 
Moreover, we expected that user engagement (H2a) and motivation (H2b) will serve as mediators in the relationship between communication style and physical activity levels, as illustrated in Fig~\ref{fig:mediator_model}. 
Therefore, we anticipated that a humorous communication style will enhance user engagement (H3a) and motivation (H3b), subsequently leading to an increase in physical activity levels (H4a+b).

To address our research questions and hypotheses, we designed a randomized controlled trial involving 66 participants divided into three groups: one with a humorous chatbot intervention, another with a non-humorous chatbot intervention as well as a no-intervention control group. The chatbot interactions are based on five BCTs, selected for their applicability in encouraging physical activity. Over a span of 10 days, participants will engage with the chatbot every other day and respond to daily survey that assess levels of engagement, motivation, and physical activity levels.
Through this study, we expect to establish a positive correlation between the use of humor in chatbot-delivered interventions and increased levels of physical activity. We anticipate that the incorporation of humor will not only enhance user engagement and motivation but will also lead to a subsequent increase in physical activity. By investigating these aspects, this study aims to contribute a nuanced understanding of the potential for humor as a viable strategy in chatbot-mediated health interventions, thereby offering empirical support for the development of more engaging and effective interventions for the intelligent conversational agents or interfaces. 


\section{Related Work}

\subsection{Behavioral Change Techniques and Humorous Communication Style in Behavioral Intervention}

The escalating prevalence of noncommunicable diseases (NCDs) like heart disease, type 2 diabetes, and specific cancers, is a critical public health issue \cite{cecchini2010tackling,lee2017running}. Responsible for about 70\% of global deaths, NCDs highlight the urgent need for effective prevention and intervention measures. Physical activity stands out as a key modifiable lifestyle factor capable of mitigating these risks, thereby emphasizing the importance of behavioral change interventions. In tackling these challenges, health professionals commonly utilize behavioral Change Techniques (BCTs) to enhance clients' physical activity levels \cite{michie2012identification,samdal2017effective}. Originating from the work of Susan Michie and colleagues, BCTs are designed as actionable strategies rooted in distinct psychological theories and empirical evidence. These techniques influence an individual's cognition, emotion, and behavioral towards achieving sustainable lifestyle changes such as regular exercise \cite{dombrowski2012identifying, michie2009effective, michie2012theories}.
Several core components often make BCTs effective. Firstly, they offer clear and specific goals, enhancing an individual's focus and motivation \cite{michie2012theories}. Secondly, BCTs often employ self-monitoring \cite{michie2012theories,michie2011strengthening}, where individuals track behaviorals, thoughts, or emotions related to their targeted behavioral, increasing self-awareness and helping pinpoint triggers or patterns. Thirdly, these techniques work through Mechanisms of Action (MoAs) \cite{michie2012theories}, which are underlying psychological processes essential for driving behavioral change. One such MoA is feedback and reinforcement \cite{michie2012theories}, which boosts adherence to behavioral change by providing progress updates and rewarding positive behaviorals. Lastly, cognitive restructuring is incorporated to challenge and shift obstructive thoughts or attitudes, thereby facilitating lasting behavioral modification. By leveraging these components, BCTs offer a nuanced yet practical approach to promote physical activity, serving as a vital tool in the ongoing fight against the global NCD crisis.

However, effective health interventions are not just about concrete strategies but also hinge on a strong therapeutic alliance between counselors and clients \cite{greenberg2003therapeutic, horvath1993role, ardito2011therapeutic}. This alliance is characterized by mutual trust \cite{ardito2011therapeutic}, shared goals, and open communication, serving as a foundational aspect of successful therapy. 
A notable element often infused into this relationship is humor. Healthcare professionals increasingly employ humor to enhance message efficacy and strengthen the therapeutic alliance \cite{blanc2014humor}. It serves as a versatile communication tool that not only lightens the conversational atmosphere but also increases client engagement and receptivity to health advice \cite{sultanoff2013integrating}. The positive effects extend to heightened motivation, reduced stress, and enhanced commitment to behavioral change \cite{blanc2014humor, panichelli2018humor, stiwi2022efficacy}. The utility of humor, however, is nuanced and influenced by cultural \cite{jiang2019cultural} and age-related factors \cite{stanley2014age}. Different cultures and generations have specific humor preferences rooted in their unique norms, values, and experiences. Therefore, understanding the type of humor used is vital. For example, affiliative humor, characterized by witty banter and jokes that promote social connection, is generally well-received across different cultures and age groups. This is in contrast to other humor types like aggressive or self-defeating humor, which can be divisive or emotionally draining.
Research underscores the effectiveness of affiliative humor in health counseling \cite{dziegielewski2003humor}, as it fosters a comfortable, trusting relationship, thereby boosting client engagement and motivation. Therefore, the strategic use of humor, particularly the affiliative type, can serve as a potent adjunct to traditional behavioral change techniques in health interventions \cite{hussong2021use}.


\subsection{Humor in Chatbot-delivered Intervention for Behavioral Change}

Given the limitations in the availability and affordability of healthcare professionals, there's a growing demand for cost-effective alternatives for promoting physical activity \cite{bertram2019using}. Conversational agents, or chatbots, have emerged as a promising solution to deliver personalized, budget-friendly interventions at scale \cite{vandelanotte2016past, zhang2018advantages}. Chatbots allow for delivering tailored and cost-efficient programs to prevent diseases and promote healthy behavioral change through natural language conversations \cite{laranjo2018conversational}. Pre-scripted as well as generative chatbots are commonly used in behavioral change interventions \cite{bickmore2013randomized,kocaballi2019personalization}. While pre-scripted chatbots provide many users with a consistent and standardized intervention, generative chatbots tailor their responses to individual user needs and preferences. Nevertheless, pre-scripted chatbots are commonly used in research studies as they offer better control over intervention content than adaptive chatbots \cite{bickmore2013randomized,kocaballi2019personalization}. Researchers can design pre-scripted responses and conversation flow for consistent delivery of behavioral change messages, allowing for easier replication of interventions and facilitating comparison across research settings.

Like human counselors, chatbots can incorporate BCTs to effectively promote physical activity \cite{wlasak2023supporting}. Their efficacy in doing so, however, is influenced by key mediating factors such as user engagement and motivation \cite{oosterveen2017systematic,teixeira2012exercise}. While chatbots can deliver content effectively, they often lack the "human touch," which is critical for maintaining long-term engagement and motivation. 
Interestingly, recent research suggests that incorporating humor into chatbot-based interventions could bridge this gap. A study by Olafsson et al. \cite{olafsson2020motivating} found that users were more engaged and motivated when interacting with a humorous chatbot compared to a non-humorous one. This led to higher likelihoods of sustained interaction and intervention adherence. Thus, incorporating humor into chatbot interventions can address the limitations of lacking human touch, empathy, and personal connection \cite{kim2012anthropomorphism, pfeuffer2019anthropomorphic}. Chatbots may bridge this gap and create a more engaging and motivating environment for users by providing a light-hearted and humorous communication style, making users more motivated to remain committed to making positive changes in their physical activity levels \cite{okun2014effective,shao2023empathetic}. While humor alone does not lead to physical activity change, it can facilitate user engagement and motivation, which are essential for actual behavioral change \cite{glanz2015theory}.

However, existing research has only measured these outcomes in isolated interactions, leaving a gap in understanding how a humorous chatbot impacts long-term user engagement and motivation for physical activity change. Besides, the humor implemented the chatbot was pre-scripted separate jokes, rather than implemented through the entire dialogue. Our study aims to fill these voids by examining the effects of a humorous communication style personalized by LLM (i.e., ChatGPT \cite{chatgpt}) in a 10-day chatbot intervention on user engagement, motivation, and physical activity levels.


\section{Study Method}

\begin{figure}[htbp]
\centering
\includegraphics[width=0.78\textwidth]{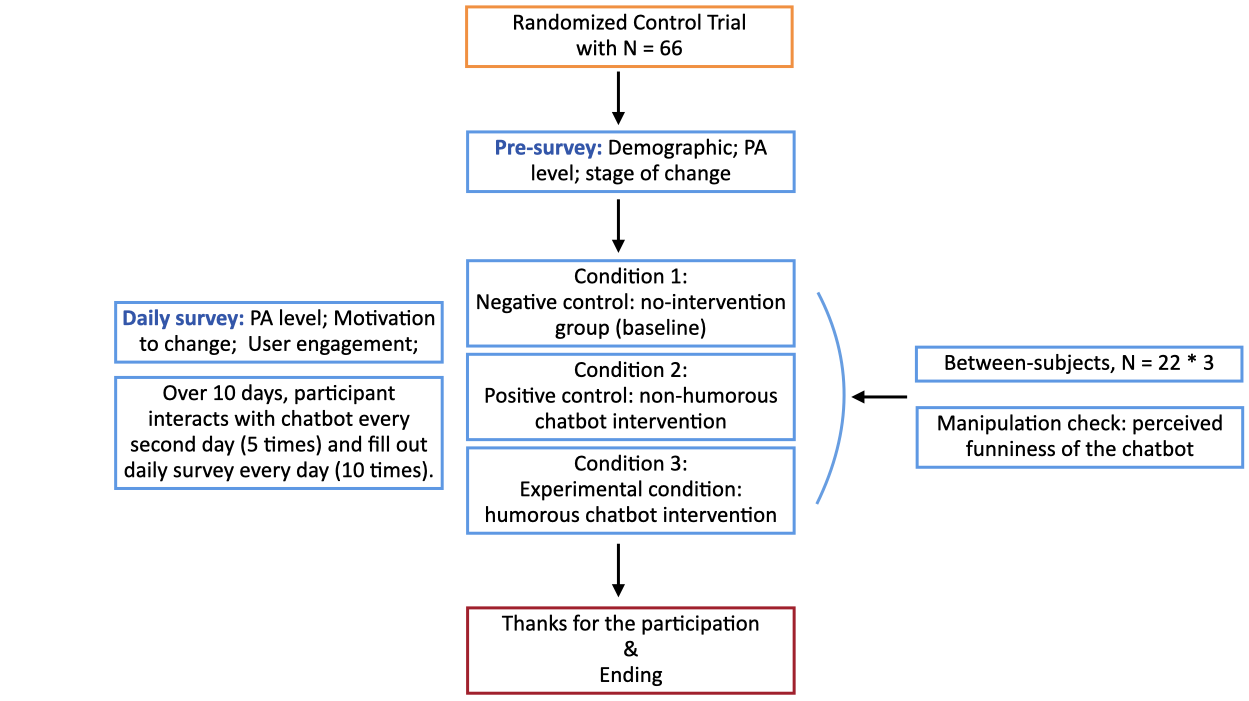}
\caption{The Overview of the Study Design.}
\label{fig:study design}
\end{figure}

The study followed a randomized controlled trial (RCT) design, investigating the effect of chatbot communication styles on a physical activity intervention. Participants were randomly assigned to one of three groups: an experimental group (humorous chatbot), a positive control group (non-humorous chatbot), and a negative control group (no chatbot intervention). This setup allowed for evaluating the impact of humorous communication while maintaining consistent intervention content for the experimental and positive control groups.
The positive control group offered a reference for assessing the humorous chatbot’s effectiveness, while the negative control group provided a baseline, showcasing natural variability in physical activity levels without any intervention. 
Comparisons across these groups would help attribute observed effects to the intervention components or the communication style manipulation. Fig~\ref{fig:study design} gives the overview of the study design.


\subsection{Participants}

The a-priori power analysis using G*Power Version 3.1 \cite{faul2007gpower} determines a required sample size in each experimental group to achieve 80\% power. This would detect a small to medium effect (f=0.15) with a significance level of $a$=.007 (adjusted for multiple testing). This calculation considered a multilevel time series analysis with ten measurements and a correlation among repeated measures of $r$=0.5. The predetermined effect size, based on \cite{olafsson2020motivating}, and the final sample conformed to these requirements.
We recruited 66 participants (N=66) through the institute’s study recruitment system. Eligibility criteria included being at least 18 years old, English proficiency, reliable internet access, and possession of a functional phone to use Telegram \cite{telegram} and the PIEL survey app \cite{piel_survey}. Participants received compensation from the institute. The study was approved by the Ethics Review Board of the institute. 


\subsection{Materials}

\subsubsection{Conversational Design with BCTs for Physical Activity Intervention}
We designed the chatbot conversations on diagrams.net~\cite{drawioDrawio}. The design involved developing a dialogue flow constructed to encourage healthy physical activity by applying five different behavioral change techniques (BCTs), i.e., the Goal Setting, Implementation Intentions, Social Environment, Cognitive Restructuring, and Small Wins. These BCTs were implemented in the chatbot conversations to create a comprehensive, evidence-based approach to promoting physical activity~\cite{}. We selected these five BCTs for their ability to create a structured, supportive, and motivating environment within a short timeframe. While long-term behavior change requires sustained efforts~\cite{prochaska1997transtheoretical}, the selected BCTs intended to establish a foundation for behavior change within our 10-day study.

\begin{figure}[ht!]
\centering
\includegraphics[width=0.88\textwidth]{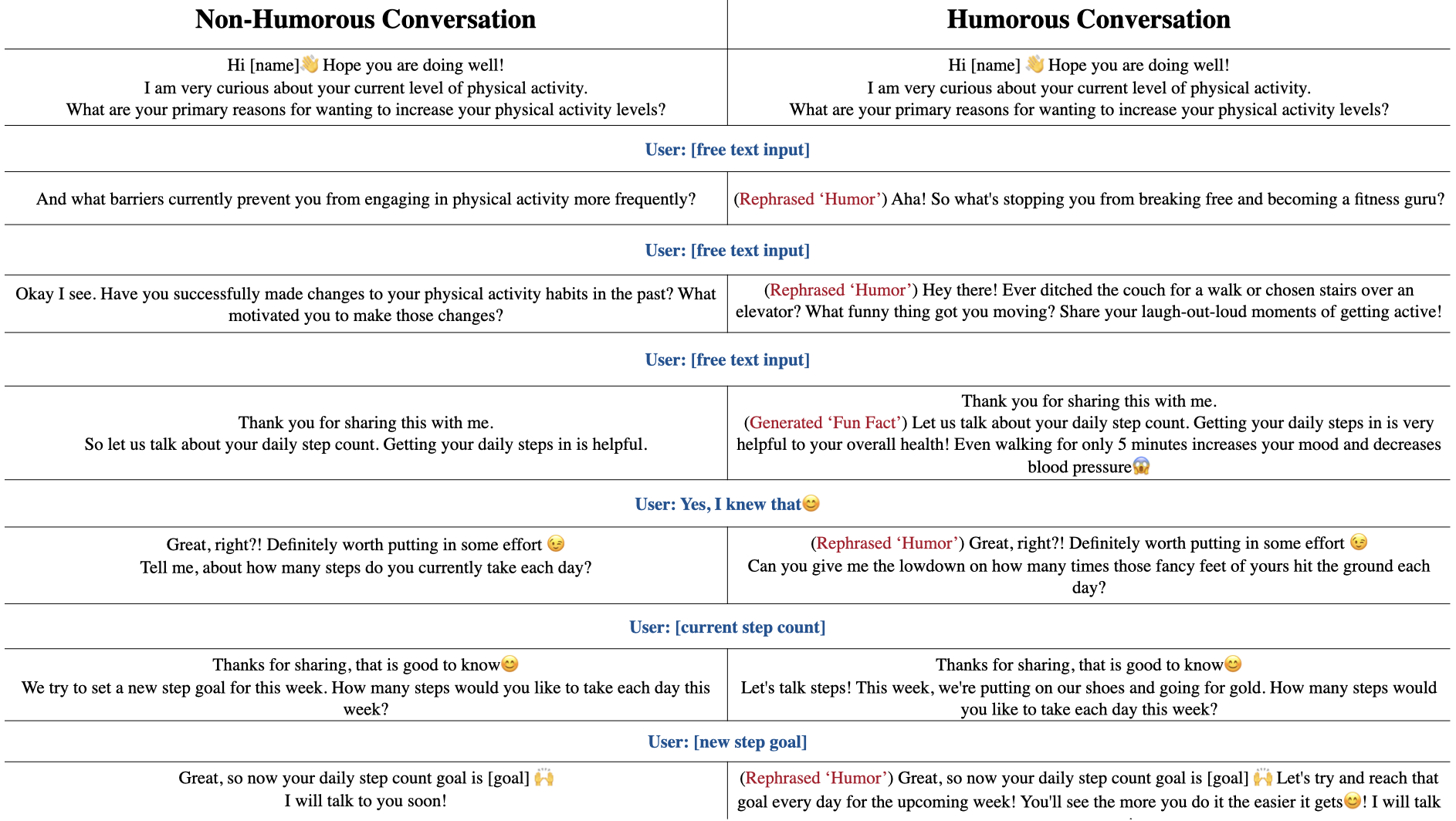}
\caption{An example of the conversation based on BCT "Goal Setting" with Non-humorous vs. Humorous communication style.}
\label{fig:conversation}
\end{figure}

We incorporated decision-making points throughout the dialogue flow to allow user choices within the chatbot conversations. At these points, the chatbot gave the users multiple predefined options to select their preferred actions or responses. An example of conversational flow is illustrated in Fig~\ref{fig:conversation}. For instance, users could choose to receive additional explanations and examples for theories or indicate if they were already familiar with the concept. These choices allowed users to customize their experience and receive tailored recommendations based on their preferences. In addition to providing predefined options, the chatbot was designed to accept free text input from users at predetermined points in the conversation. This feature facilitated an interactive and personalized experience, allowing users to communicate in their own words, and provided the input for LLM (i.e., ChatGPT) to generate more humorous dialogue. For instance, when the chatbot inquired about physical activity preferences, users could provide specific details such as their preferred types of exercise, time of day, or locations for physical activity. 
This free-text input feature was included primarily to create a more natural interaction between the user and the chatbot.

As shown in Fig \ref{fig:chatbot} and Fig \ref{fig:conversation}, the humorous and non-humorous conversations shared the same basic dialogue content and length but differed in the style of the chatbot responses. The non-humorous conversations were planned and scripted first, with a neutral communication style, and then served as the basis for the development of the humorous condition. In creating the humorous condition, the tone of the conversation was deliberately adjusted at appropriate junctures. The goal was to introduce humor analogous to how human health professionals might apply it in real-life interactions. It was essential to strike a balance wherein humor enhanced user engagement without overshadowing the primary objectives of the conversation, i.e., conveying the BCTs and promoting behavior change. 
The intention was not to make the conversations hilariously entertaining but to employ humor strategically to augment user interest while remaining professional.


To ensure the clarity and flow of all conversations and the effectiveness of humor, we followed a think-aloud protocol with three participants who were not involved in the subsequent experiment. A think-aloud protocol is a qualitative research method where participants verbalize their thoughts while performing a task, offering insights into cognitive processes and user experiences \cite{alhadreti2018rethinking}. It is widely used in usability testing, user experience (UX) research, cognitive psychology, and human-computer interaction (HCI) studies to optimize user interfaces and experiences \cite{guss2018thinking,joe2015use}. For the think-aloud protocol, we followed these steps: 1) Reading the conversation script to the participants. 2) Discussing their perceptions, understanding, and any issues they encountered concurrently, aiming to identify and solve misunderstandings and confusion. 2a) For clarity of content (BCTs) and conversation flow, we focused on whether explanations for the conversation content are straightforward and transitions between topics flow naturally. 2b) The effectiveness of humor was tested by assessing participants' reactions to funny instances. We ensured that jokes were never perceived as offensive but as positive, inclusive, light-hearted, and friendly (affiliative). 3) We incorporated the feedback from participants and iterated the process until we achieved satisfactory results.


\subsubsection{Technical Implementation of Chatbot and Humors}

The chatbots, humorous and non-humorous alike, were developed using the RASA framework~\cite{rasa}.
The chatbot was built upon three core components: \\
\textbf{1). Natural Language Understanding (NLU) Component:} This vital element interprets user inputs and interactions, ensuring the accurate understanding of messages and intentions conveyed by participants, including responses to interactive buttons.\\
\textbf{2). Dialogue Management Component:} This module predicts the chatbot’s responses within conversations. Utilizing the history of ongoing dialogues, it provides contextually appropriate and coherent replies.\\
\textbf{3). Dialogue Data Component:} This element encompasses the pre-scripted BCT-grounded conversations fed into the dialogue management module. Together, these modules enabled a smooth, engaging interaction with participants, with the chatbot understanding inputs through the NLU, generating responses via the dialogue management module, and relying on pre-scripted content provided by the Dialogue Data component. 

\begin{figure}[ht!]
\centering
\includegraphics[width=0.725\textwidth]{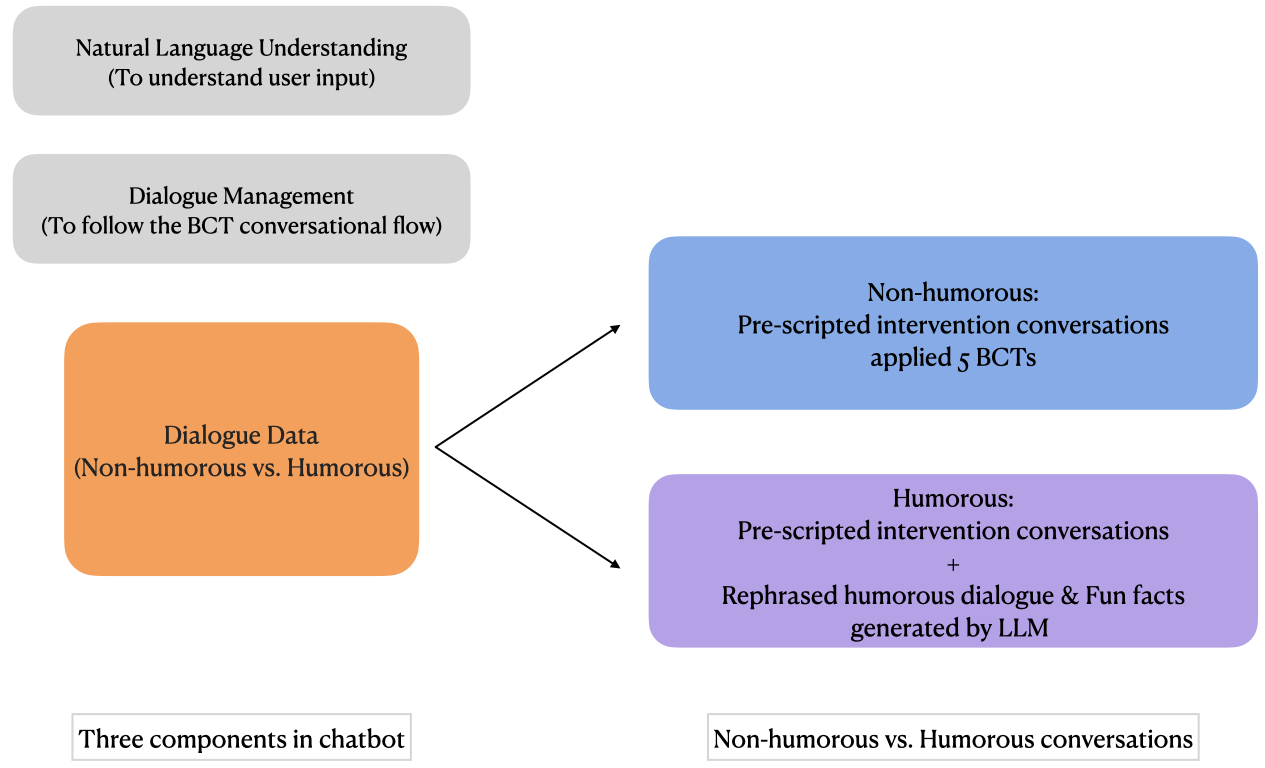}
\caption{Components in the chatbot and the approach utilized to implement the Non-humorous vs. Humorous communication style in the chatbot.}
\label{fig:chatbot}
\end{figure}

In addition to the implementation of the BCT-grounded conversations, for the humorous communication style, we worked with the generative large language model (LLM), i.e., ChatGPT~\cite{chatgpt}, to make the communication style of the chatbot more humorous. 
For the humorous chatbot, we used the Dialogue Data component as in the non-humorous counterpart but incorporated the generated humorous conversations. These were crafted by leveraging an LLM (i.e., GPT-4 model) to transform pre-scripted non-humorous dialogues into expressions with a more socially expressive and affiliative humorous style. 
Moreover, we can generate personalized fun facts derived from dialogue history and user profiles by LLM. For example, if a dialogue prompted users to discuss reasons and barriers for behavioral change related to physical activity, the chatbot could utilize the user input to craft personalized fun facts. 
Fig \ref{fig:chatbot} demonstrates this process: non-humorous dialogues, associated with specific BCT, were transformed into humorous expressions, with fun facts generated based on dialogue history.
Applying LLM allowed us to tailor the chatbot’s responses by implementing affiliative humor. Therefore, the humorous segments aimed to create a positive and enjoyable social environment and foster a sense of camaraderie and friendliness between the chatbot and the user. 
For instance, the non-humorous version would say, "Why do you think you did not reach your goal? Was the number of steps a bit too much for now? Or was there something else that held you back?" While the humorous version would say, "Why the step shortage? Did you get caught up in a Netflix binge, or were your feet on strike demanding higher wages?" 
A comparison of (non-) humorous chatbot conversations can be seen in Fig \ref{fig:conversation}.

Upon completion of the chatbot's development, we engaged in extensive testing and validation to ensure adherence to the pre-scripted flow of BCTs and accurate delivery of motivational content. Following successful tests, the chatbot was deployed on Telegram, chosen for its accessible and user-friendly interface, facilitating easy participant engagement with the chatbots.


\subsubsection{Measures}
Throughout the 10-day study, ecological momentary assessment (EMA) data was collected. EMA data encompasses real-time, momentary data obtained from individuals in their daily lives and natural environments, mitigating reliance on memory or general impressions of behavior~\cite{shiffman2008ecological}. This method facilitates the assessment of user engagement, motivation, and physical activity levels over time. Participants were instructed to install the PIEL Survey App~\cite{piel_survey} on their mobile phones to facilitate EMA data collection. 
The data collection process was segmented into two stages: an initial pre-survey on the first day and a daily survey for the subsequent ten days to measure the participant's engagement level, motivational level, and daily physical activity as well as the manipulation check for the funniness of the chatbot.

\paragraph{Pre-survey.} Initially, participants provided socio-demographic information (including age, gender, home country, education level, and first language) and their current physical activity levels. Pre-intervention physical activity was determined using the recorded step counts from participants' mobile phones from the previous week, with the total step count for each day being utilized. This measurement methodology aligns with research previously conducted by Motl et al~\cite{motl2012reactivity,motl2017randomized} and Pilutti et al~\cite{pilutti2014randomized}.

\paragraph{User Engagement.} Each day at 9 pm throughout the 10-day study duration, data about user engagement, motivation, and physical activity levels were collected. 
User engagement during interactions with the chatbot was measured using the engagement questionnaire from Olaffson et al~\cite{olafsson2020motivating} study. This questionnaire comprises ten items, each answered on a 5-point Likert scale ranging from "strongly disagree" to "strongly agree", with sample items including "I would like to continue working with the chatbot" and "I would recommend the chatbot to others". Demonstrating solid psychometric properties, the questionnaire boasts high internal consistency ($a$ = 0.85) and test-retest reliability ($r$ = 0.72).

\paragraph{The Motivation to Change Physical Activity Questionnaire - Short Form (MCPAQ-SF).} Participants' motivation to alter their physical activity habits was assessed using the MCPAQ-SF \cite{mcauley1992psychometric}, which includes 11 items, each rated on a 5-point Likert scale. Sample items encompass statements like "I am confident I can successfully incorporate regular physical activity into my lifestyle" and "I am willing to adjust my daily routine to accommodate regular physical activity". The MCPAQ-SF exhibits sound psychometric properties, including high internal consistency ($a$ = 0.78 to 0.88), test-retest reliability ($r$ = 0.73 to 0.85), and significant correlations with other physical activity motivation measures, even predicting alterations in physical activity behavior over time \cite{mcauley1992psychometric}.

\paragraph{Daily Physical Activity (Step Count).} Physical activity levels were monitored daily using participants’ step counts from their mobile phone trackers.

\paragraph{Manipulation Check.} To assess participants' perception of the chatbot's humor and gauge their appreciation levels, the statement "The chatbot is funny" was included as a manipulation check item, with responses ranging from "strongly disagree" to "strongly agree" on a 5-point scale.


\subsection{Procedure}

Following randomization, participants received general study information without details about the specific conditions to prevent biased responses. For instance, they were informed about the investigation into chatbot interactions’ effects without disclosure of the communication styles used. Participants in the negative control group were told about a study into user experiences without an intervention.
After obtaining informed consent, participants were instructed to use the PIEL Survey App \cite{piel_survey} for questionnaires and access the assigned chatbot on Telegram \cite{telegram}. Participants in the (non-) humorous chatbot groups initiated conversations by pressing the START button or typing "Hi". The chatbot would then ask participants to indicate the specific day of the conversation to start that conversational topic. Reminders were sent every second day to prompt participants to engage in another conversational interaction with the chatbot.

\begin{figure}[htbp]
\centering
\includegraphics[width=0.93\textwidth]{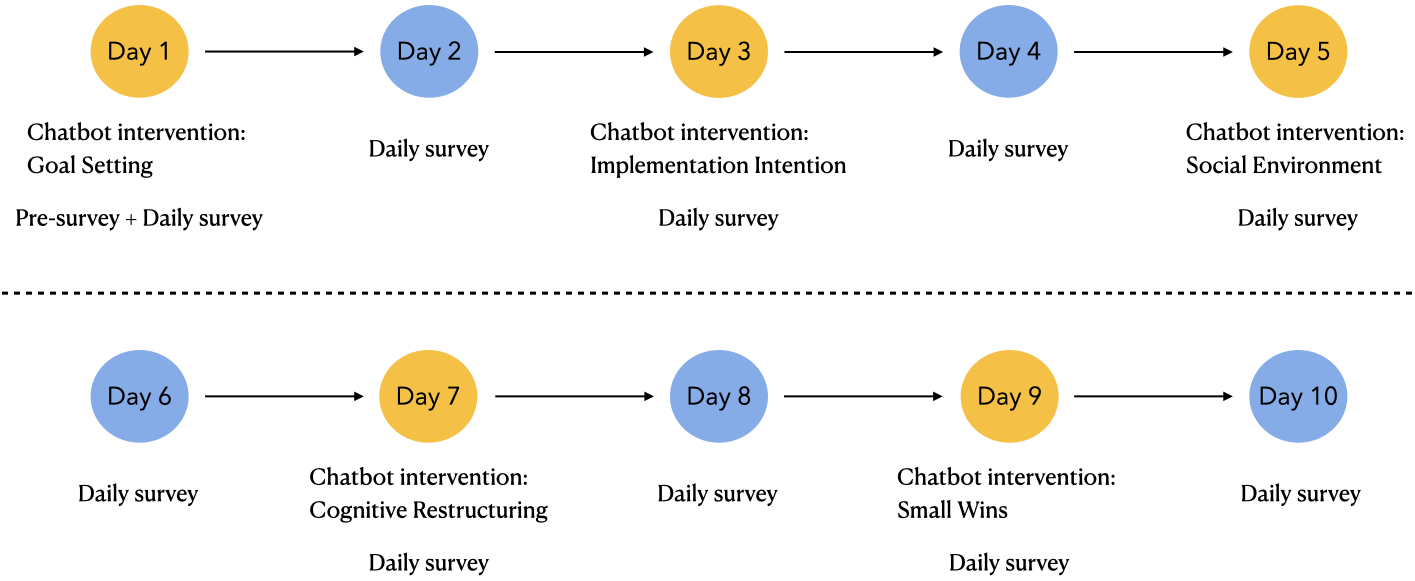}
\caption{Illustration of the 10-Day Study Progression.}
\label{fig:procedure}
\end{figure}

On the first day of the study, participants filled in the pre-survey and started the daily survey. Over ten days, participants in the chatbot groups had five conversations with their assigned chatbot, each employing a distinct BCTs for physical activity introduced in the Section 3.2.1. 
Every second day of the 10-day study, the chatbot initiated different BCT conversations with the user. A flowchart of the study procedure is shown in Fig \ref{fig:procedure}. Participants in the no-intervention group only completed questionnaires related to motivation and physical activity levels, but no user engagement with the chatbot involved.


\subsection{Data Analysis}

Participants’ engagement and motivation scores were computed by summing responses across items for each day, resulting in ten separate scores per participant.
We first checked the assumptions underlying our statistical models. The normality of residuals was confirmed through density and Q-Q plots, and multicollinearity was assessed through a correlation matrix, revealing no substantial correlations among predictors. While slight violations of linearity and homogeneity of variance assumptions were noted, these were considered manageable due to the flexibility of multilevel models in capturing trends and accommodating variations in time series data.
For handling missing data, we employed the missRanger package in R to utilize a Random Forest-based single imputation approach \cite{mayer2019packagemissranger}, which allows for sophisticated pattern-based value prediction. Hypothesis testing was conducted using multilevel time series analyses through the lme4 \cite{bates2015fitting} and lmerTest \cite{kuznetsova2017lmertest} packages in R, supplemented with multilevel model ANOVA via the car package \cite{fox2011companion}. Sum-to-zero contrasts were established for facilitating ANOVA and post hoc tests.

The analyses were conducted separately for each hypothesis and the manipulation check, involving distinct sets of predictor and outcome variables. 
The manipulation check assessed whether the humorous chatbot was perceived as funnier than the non-humorous one to validate our manipulation's effectiveness, with condition (n=3) and intervention day (n=10) as fixed effects, with intervention day nested within the condition, and included participant id (N=66) as a random effect. 
Hypothesis 1 investigated the effect of a (non-)humorous chatbot communication style on physical activity levels over time, considering condition (n=3) and intervention day (n=10) as fixed effects, with intervention day nested within condition, and included participant id (N=66) as a random effect. 
Additionally, we investigated potential differences in pre-intervention physical activity compared to activity levels during the intervention for each condition. This analysis was carried out separately for each condition using multilevel time series analyses with time point (n=2, pre-intervention and intervention) and experiment day (n=17, with seven no-intervention days and ten intervention days) as fixed effects and day nested within time point. Participant id (N=66) was considered as a random effect.
For hypothesis 2, we used the mediation package \cite{tingley2014mediation} to explore how user engagement and motivation mediate the relationship between communication style and physical activity levels. The condition was indicated as the treatment variable, while engagement (H2a) and motivation (H2b) were set as the mediator, respectively. We used the default number of simulations (sims=1000). 
Hypothesis 3 anticipated higher engagement (H3a) and motivation (H3b) levels with a humorous communication style. In the analyses for H3, we considered condition (n=3) and intervention day (n=10) as fixed effects, with intervention day nested within the condition, and included participant id (N=66) as a random effect. 
Lastly, hypothesis 4 explored the relationship between user engagement (H4a), motivation (H4b), and physical activity levels.
The significance of effects was evaluated through ANOVA, with post hoc pairwise comparisons conducted using the multcomp \cite{hothorn2008simultaneous} and lsmeans \cite{lenth2016leastsquares} libraries, applying Holm’s method for multiple comparisons. All models controlled for participants’ age and gender.


\section{Findings}


\subsection{Participants Characteristics}
The study comprised 66 participants aged between 18 and 65 years (M=24, SD=7.46), including 18 males, 46 females, and two non-binary individuals. Participants' education levels varied: 28 had completed high school, 29 held undergraduate degrees, and nine had postgraduate qualifications. While most participants (n=26) reported German as their first language, others listed Dutch (n=20), English (n=9), Croatian (n=4), Slovenian (n=4), or Latvian (n=3) as their primary language. Detailed information of the participants can be found in Table \ref{table:participants}.


\begin{table}[htbp]
\centering
\renewcommand{\arraystretch}{0.92}
\begin{tabularx}{\columnwidth}{p{5cm} >{\raggedright\arraybackslash}p{6cm} >{\raggedright\arraybackslash}X}
\toprule
\textbf{Demographic} & \textbf{Categories} & \textbf{Numbers of Participants (\%)} \\
\midrule
Gender & & 
\\ 
 & Female & 46 (69.7\%)
\\ 
 & Male & 18 (27.3\%)
\\ 
 & No-binary & 2 (3.0\%)
\\
\hline
Age & & 
\\
 & 18-65 & M=24 (SD=7.46)
\\
\hline
Education & & 
\\
 & High school & 28 (42.4\%)
\\
 & Undergraduate & 29 (43.9\%)
\\
 & Postgraduate & 9 (13.6\%)
\\
\hline
Primary language & & 
\\ 
 & German & 26 (39.4\%)
\\
 & Dutch & 20 (30.3\%)
\\
 & English & 9 (13.6\%)
\\
 & Croatian & 4 (6.1\%)
\\
 & Slovenian & 4 (6.1\%)
\\
 & Latvian & 3 (4.5\%)
\\
\bottomrule
\end{tabularx}
\caption{Characteristics of Participants}
\label{table:participants}
\end{table}


\subsection{Descriptive Statistics}
Table \ref{table:descriptive} provides a comparative overview of the average physical activity levels, engagement levels, and motivation scores across the three different conditions. Positive relationships were observed between engagement and motivation, engagement and physical activity, and engagement and the rated humor of chatbot interactions (see Table \ref{table:correlation}). Conversely, higher age was associated with reduced engagement. 
Additionally, higher motivation levels were linked to increased physical activity. 
Motivation showed no significant correlation with the funniness rating or age. 
Lastly, physical activity positively correlated with the funniness rating, while its correlation with age was not statistically significant. 
No significant correlation was found between the funniness rating and age.


\begin{table}[htbp]
\centering
\renewcommand{\arraystretch}{1.35} 
\begin{tabularx}{\columnwidth}{p{4cm} >
{\raggedright\arraybackslash}X >
{\raggedright\arraybackslash}X >
{\raggedright\arraybackslash}X}
\hline
\textbf{Variables} & \textbf{} & \textbf{Conditions} & \textbf{} \\
\hline
 & Humorous M(SD) & Non-humorous M(SD) & No-intervention M(SD) 
 \\
\hline
Physical Activity & 8348.63 (4834.93) & 6210.08 (2456.62) & 6465.87 (3808.79)
\\ 
\hline
Engagement & 35.25 (3.58) & 31.58 (5.22) & -
\\ 
\hline
Motivation & 44.49 (5.05) & 43.75 (4.74) & 43.67 (4.66)
\\
\hline
\end{tabularx}
\caption{Descriptive Analysis per Condition.}
\label{table:descriptive}
\end{table}


\subsection{Manipulation Check}
\begin{figure}[ht!]
\centering
\includegraphics[width=0.47\textwidth]{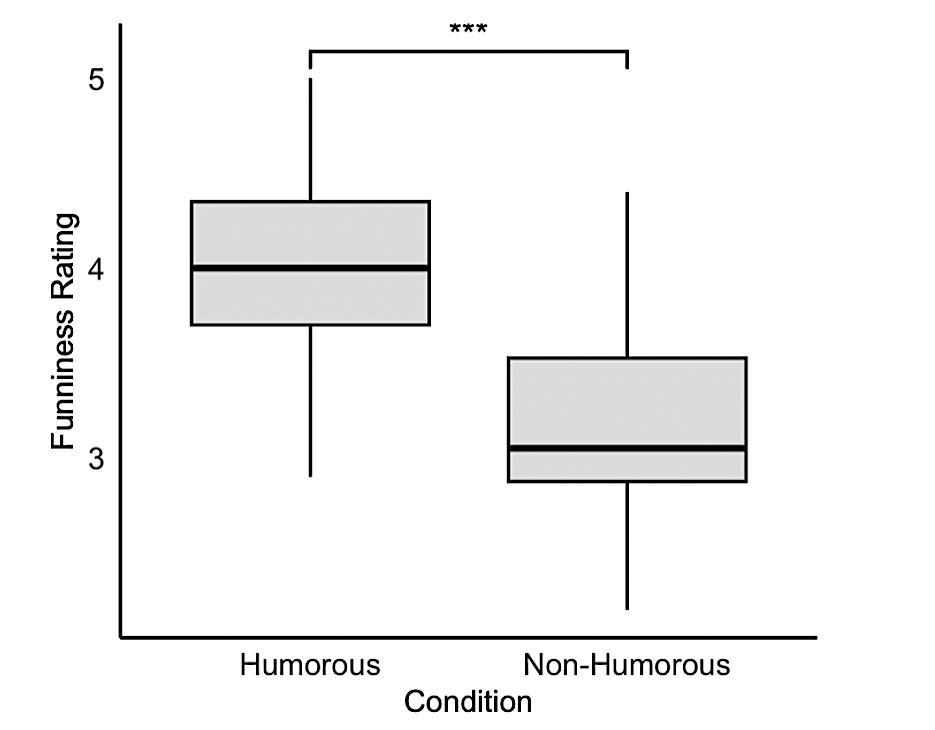}
\caption{Comparison of Mean Rated Funniness Between Conditions.}
\label{fig:funniness}
\end{figure}

Ensuring that the humorous chatbot was perceived as funnier than the non-humorous one was crucial to validate our manipulation's effectiveness. We employed a multilevel model to test this, which predicted perceived funniness based on participants' assigned conditions. The analysis revealed a significant relationship, showing that the assigned condition (humorous vs. non-humorous) strongly predicted the perceived funniness of the chatbot interaction, F(1, 40)=14.39, p < .001. This result confirmed the success of our manipulation, as the humorous condition was perceived as funnier than the non-humorous condition as illustrated in Fig \ref{fig:funniness}.


\subsection{Confirmatory Analyses}

\subsubsection{Effect of humorous communication style on physical activity levels (Hypothesis 1)}

\begin{figure}[htbp]
\centering
\includegraphics[width=0.74\textwidth]{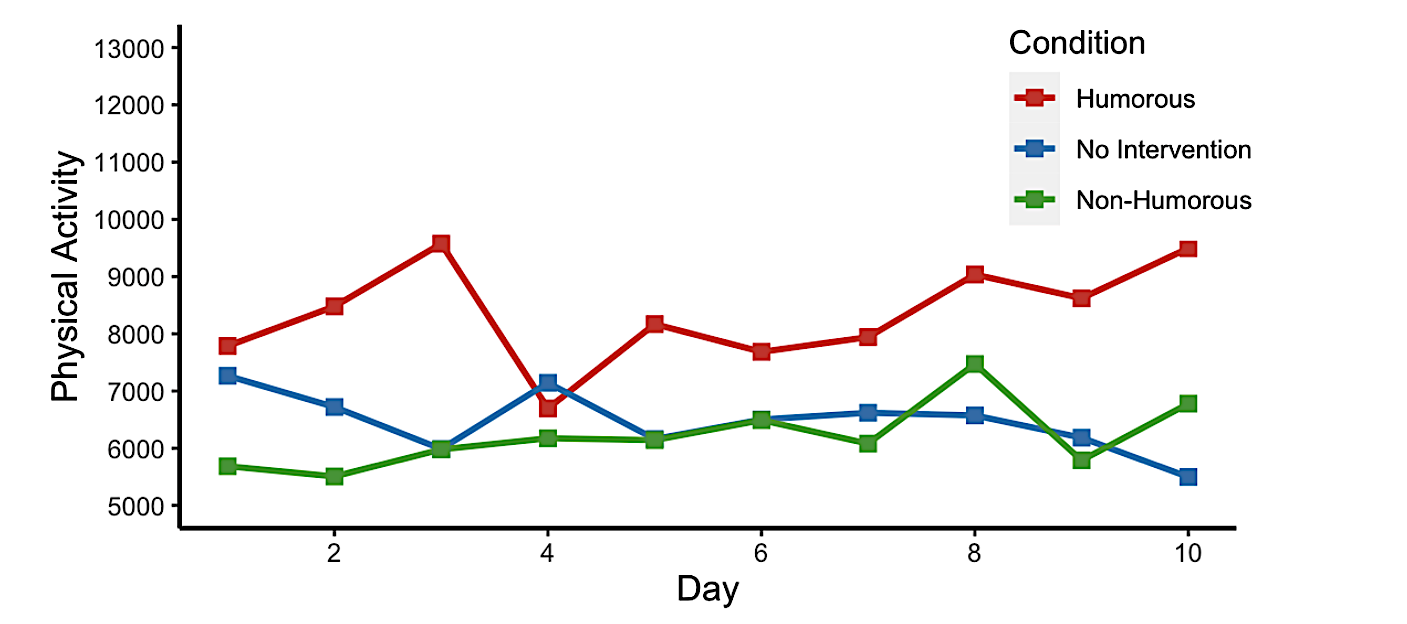}
\caption{Mean Scores of Physical Activity (daily steps) per Condition Throughout the Intervention.}
\label{fig:pa}
\end{figure}

A multilevel time series model was used to test the hypothesis that the chatbot health behavior intervention using a humorous communication style has a stronger positive effect on physical activity levels over time than a non-humorous communication style (H1), with physical activity (measured in step counts) as the outcome variable and condition (categorized as no-intervention, non-humorous, or humorous) as the predictor variable. 
Condition significantly predicted physical activity levels, F(2, 63)=5.23, p=.008. Post hoc pairwise comparisons with a Holm correction indicated that the humorous condition was associated with significantly higher physical activity levels than both the non-humorous condition, t(2178.5)=2.90, p=.016, and the no-intervention condition, t(1931.9)=2.69, p=.018 (Fig \ref{fig:pa}). Lastly, the non-humorous condition was not linked to significantly higher physical activity levels than the no-intervention condition, t(246.7)=0.36, p=.739. 
This result supports the initial hypothesis that the humorous chatbot would increase physical activity levels over time more than a non-humorous chatbot.


Similarly, we employed a multilevel time series model to explore whether there were differences in physical activity levels throughout the intervention in comparison to the indicated no-intervention activity within each respective condition, with physical activity as the outcome variable and time point (categorized in no-intervention vs. intervention) as the predictor variable. While time point significantly predicted physical activity in the humorous condition, F(1, 336)=8.35, p=.004, this was not the case for the non-humorous, F(1, 304)=0.23, p=.618, nor the no-intervention condition, F(1, 365)=2.09, p=.149. Participants in the humorous condition showed increased physical activity in the intervention compared to their baseline level. At the same time, there was no change in physical activity in the non-humorous and no-intervention conditions.


\subsubsection{Mediation role of user engagement and motivation (Hypothesis 2)}

With a mediation analysis, we assessed whether user engagement (H2a) and motivation to change (H2b) mediate the relationship between chatbot communication style and physical activity levels, with physical activity as the outcome variable, condition as the treatment variable, and engagement and motivation as the mediators, respectively. Engagement significantly mediated the relationship between chatbot communication style and physical activity, with an average causal mediation effect (ACME) of B=717.58, 95\% CI [145.10, 1508.29], p=.004. This effect appeared to be a complete mediation, as the main effect of condition became non-significant when including the mediator in the analysis, as could be seen with the average direct effect (ADE) of B=1467.59, 95\% CI [-145.35, 3012.63], p=.072. Conversely, motivation did not significantly mediate the relationship between chatbot communication style and physical activity, with an average causal mediation effect (ACME) of B=64.26, 95\% CI [-172.60, 346.08], p=.650. These findings provide partial support for our hypothesis. Specifically, while the chatbot's communication style affected participants' physical activity levels, user engagement helps to explains this connection. However, participants’ motivation to change did not seem to influence the relationship between chatbot communication style and physical activity similarly.


\begin{figure}[htbp]
\centering
\includegraphics[width=0.7\textwidth]{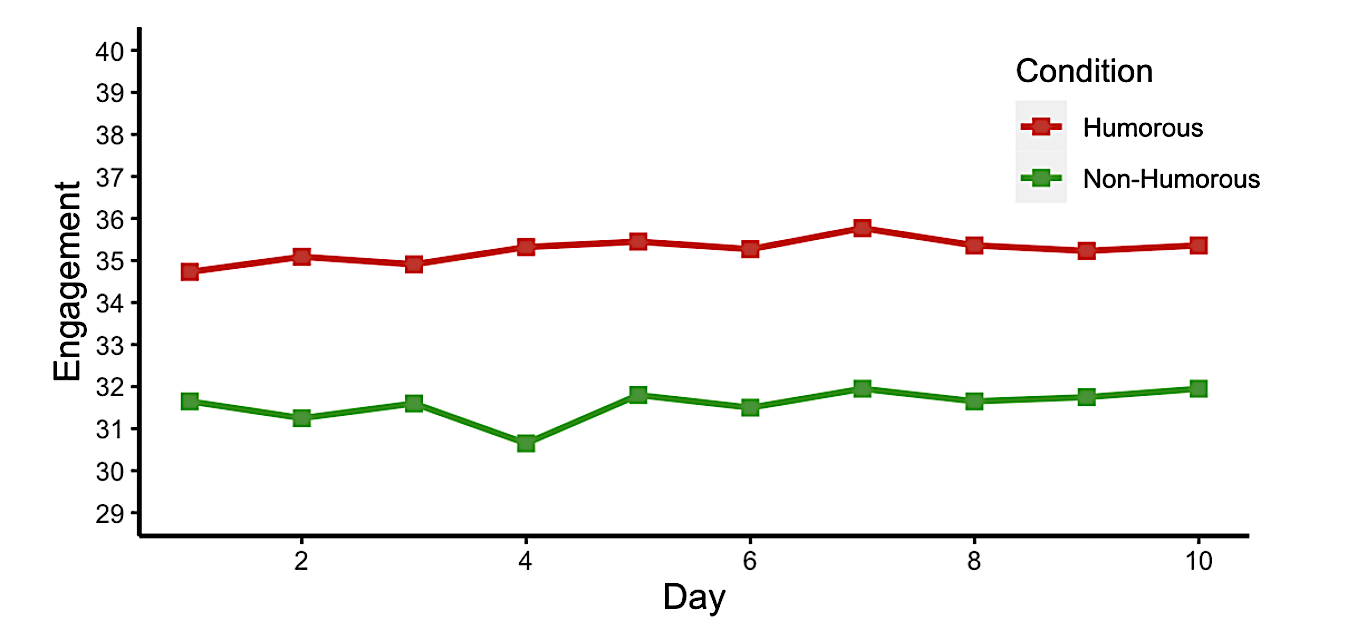}
\caption{Mean Scores of Engagement per Condition Throughout the Intervention.}
\label{fig:engagement}
\end{figure}

\begin{figure}[ht!]
\centering
\includegraphics[width=0.7\textwidth]{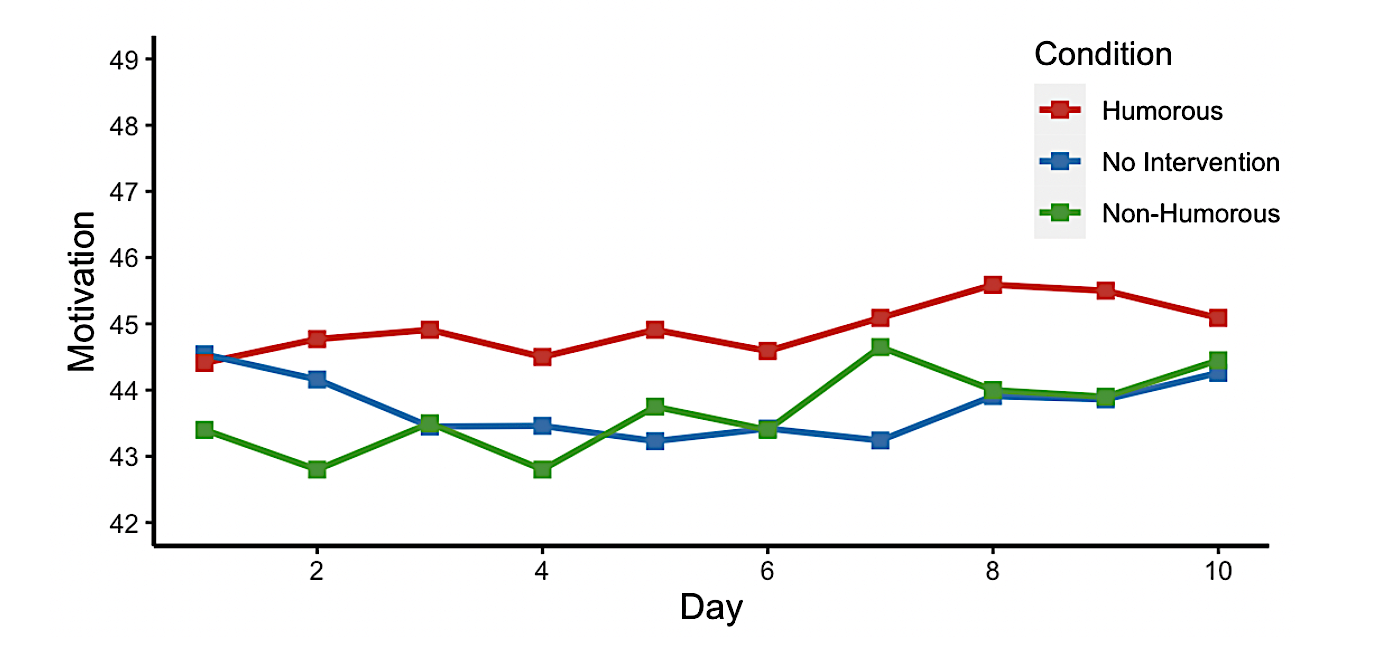}
\caption{Mean Scores of Motivation per Condition Throughout the Intervention.}
\label{fig:motivation}
\end{figure}


\subsubsection{Relationships among humorous communication style, user engagement, motivation and physical activity levels (Hypothesis 3 \& 4)}

We employed a multilevel time series model to investigate whether a humorous chatbot would increase user engagement (H3a) and motivation to change (H3b) to a greater extent than a non-humorous chatbot, with engagement and motivation as the outcome variables and condition as the predictor variable. Condition significantly predicted engagement, F(1, 53.68)=4.85, p=.032, but not motivation to change F(2, 63)=0.55, p=.577. 
These results only partly support our expectations: While the humorous chatbot effectively enhanced user engagement compared to the non-humorous chatbot (Fig \ref{fig:engagement}), there were no differences in motivating participants to change their physical activity levels across all three conditions (Fig \ref{fig:motivation}).

Besides, a multilevel time series model was used to test the hypothesis that user engagement (H4a) and motivation (H4b) increase physical activity levels, with physical activity as the outcome variable and engagement and motivation as the predictor variables, respectively. Engagement significantly predicted physical activity levels, F(1, 151.60)=7.33, p=.008, as well as motivation, F(1, 282.99)=16.21, p<.001. These results support the initial hypothesis that higher user engagement and motivation increase physical activity levels.

Ultimately, positive relationships were observed between engagement and motivation, engagement and physical activity, and engagement and the rated humor of chatbot interactions (see Table \ref{table:correlation}). 


\begin{table}[htbp]
\centering
\renewcommand{\arraystretch}{1.3} 
\begin{tabularx}{\columnwidth}{p{5.5cm} >
{\raggedright\arraybackslash}X >
{\raggedright\arraybackslash}X >
{\raggedright\arraybackslash}X >
{\raggedright\arraybackslash}X}
\hline
\textbf{Variables} & \textbf{1} & \textbf{2} & \textbf{3} & \textbf{4} \\
\hline
Engagement &  &  & 
\\ 
\hline
Motivation & .49 &  & 
\\
\hline
Physical Activity & .34 & .28 & 
\\
\hline
Funniness & .59 & .22 & .36
\\
\hline
Age & -.35 & -.13 & -.02 & -.21 
\\
\hline
\end{tabularx}
\caption{Pearson Correlation among Measured Variables.}
\label{table:correlation}
\end{table}



\section{Discussion}


This study investigated the influence of humor in a chatbot's communication style in a 10-day physical activity intervention, with a focus on its impact on user engagement, motivation, and overall levels of physical activity. Our primary objective was to determine if a chatbot employing a humorous communication style would be more effective in enhancing physical activity levels than one using a non-humorous style. Additionally, we analyzed whether user engagement and motivation acted as mediators in the relationship between the chatbot’s communication style (either humorous or non-humorous) and observed changes in physical activity levels among participants.



\subsection{The Benefit of Humorous Communication Style for Physical Activity Intervention}

Our study found that participants exposed to a chatbot with a humorous communication style exhibited increased physical activity levels compared to those interacting with non-humorous chatbots and the no-intervention group. The humorous group showcased a notable increase in physical activity levels from their pre-intervention measurements. In contrast, no significant difference in activity levels was observed between the non-humorous and no-intervention groups, suggesting the non-humorous communication style did not effectively promote physical activity. Thus, regarding the primary research question, a humorous communication style yielded the anticipated positive effect on physical activity levels. However, the non-humorous intervention failed to exert any noticeable influence on physical activity levels, deviating from our initial expectations.

The disparity in outcomes between the humorous and non-humorous interventions indicates the importance of user engagement in the success of chatbot-facilitated physical activity interventions. The increased user engagement seen in the humorous group positively correlated with higher physical activity levels, affirming the role of engagement in successful intervention outcomes. On the other hand, the lower engagement levels in the non-humorous group mirrored their lack of progress in physical activity, highlighting that without effective engagement, interventions are less likely to succeed in promoting physical activity.

Our findings align with previous research emphasizing the importance of humor as an effective communication strategy in both human and chatbot interactions in healthcare interventions \cite{olafsson2020motivating,panichelli2018humor,sultanoff2013integrating}. Humor fosters a positive, enjoyable atmosphere during interactions, encouraging patients to actively engage in their healthcare process. It has been identified as a facilitator of therapeutic alliances, promoting client engagement and receptivity to health messages, thereby establishing an atmosphere conducive to successful healthcare interventions \cite{blanc2014humor}. 
Furthermore, our study highlight the efficacy of humor in enhancing engagement within chatbot-based interventions, thereby addressing common engagement-related challenges observed in such interventions \cite{dziegielewski2003humor,glanz2015theory,olafsson2020motivating}. This mirrors prior research on human health professional counseling where humor plays a crucial role. Just as in interactions with healthcare professionals, engaging communication within chatbot interventions is essential and can be effectively achieved through the strategic use of humor \cite{kim2012anthropomorphism,panichelli2018humor,pfeuffer2019anthropomorphic}. This improved engagement, facilitated by humor, subsequently fosters tangible behavioral changes over the intervention's duration.

In sum, our findings highlight the significant potential of using humor, particularly affiliative humor, as a strategy to enhance the effectiveness and appeal of chatbot-delivered interventions, offering a viable solution to enhance user engagement and promote behavioral change in the realm of digital health interventions.


\subsection{The Role of Individual Motivation for Physical Activity Intervention}

Surprisingly, our initial expectations regarding the mediating role of motivation were contradicted. Participants' motivation did not influence the link between the chatbot's communication style and physical activity levels. 
While participants' motivation predicted their level of physical activity, the chatbot's communication style did not impact participants' motivation. 
Our study's results indicate that not only the communication style but also the overall intervention, spanning both intervention groups, did not lead to significant changes in participants' motivation levels compared to the no-intervention group: neither of the intervention groups demonstrated substantial effects on participants' motivation levels when compared to those who were not exposed to an intervention.
This stands in contrast to prior research demonstrating the efficacy of incorporating BCTs in chatbot interventions to motivate increased physical activity \cite{wlasak2023supporting}. Further, our research built on the practical foundations laid by previous studies, specifically focusing on BCTs that had demonstrated success in enhancing user motivation. 
The lack of impact on motivation might be due to the study's participant sampling and their diverse reasons for participating. Participants included in the study were drawn from convenience sources, including friends, peers, and university students, which might not accurately represent individuals who actively seek to modify their physical activity levels. 

In this context, intrinsic motivation becomes relevant. Intrinsic motivation refers to the internal desire or drive to engage in an activity for the inherent satisfaction, enjoyment, or personal interest it brings rather than being solely motivated by external rewards or pressures \cite{deyoung1985encouraging,seifert2012enhancing}. When someone is intrinsically motivated, they find the activity fulfilling and gratifying. 
In the context of behavior change, intrinsic motivation plays a crucial role: when intrinsically motivated to change a behavior, individuals are more likely to engage in the change process willingly and persistently \cite{seifert2012enhancing}. They view the behavior change as meaningful, aligning with their values, goals, and interests. The presence or absence of intrinsic motivation can significantly influence how individuals approach and engage with interventions to change their behavior. 
In contrast, if motivation is primarily driven by external factors like rewards or social pressures, it is known as extrinsic motivation \cite{dacey2008older,matsumoto2004motivational}. 
Participants in this study might not have been genuinely interested or self-motivated to alter their activity levels, which could have affected their responsiveness to the intervention. This misalignment between the intervention's objectives and participants' inherent motivations could have contributed to the observed lack of impact on motivation.


\section{Limitations and Future Work}

Our research possesses certain limitations that warrant acknowledgment. 
Firstly, the participant sample is primarily student recruited from the institute. While this provides a diverse range of motivations for participation, it may not accurately mirror the profile of individuals actively attempting to modify their physical activity. 
Secondly, the generalizability of findings is limited, considering the participant demographic was predominantly European individuals in their twenties. Though it allows for a uniform cultural, age, and lifestyle background, the data's applicability to broader populations may be restricted. 
Additionally, we recognize the effects of humor may exhibit variations across different demographic sectors and levels of technological proficiency, factors that were not central considerations in this study. 
Finally, the humor in conversations, even with the assistance of large language models like ChatGPT, was pre-defined, potentially limiting the organic and dynamic insertion and effectiveness of humor within the dialogues.

Addressing these limitations, future work offers promising avenues for exploration and refinement. 
Researchers might consider engaging participants who are at different stages of physical activity modification to obtain a representation closer to target populations actively engaging with behavioral change. There is also an opportunity for future studies to explore the influence and effectiveness of BCTs within various demographic groups, encompassing diverse cultural, age, and lifestyle backgrounds. This approach would enable a more nuanced understanding and application of humor-infused BCTs globally. 
Furthermore, subsequent studies should consider the individual differences in humor effects and tech-savviness as potential modulators of humor's impact on motivation, offering more personalized and effective interventions. 
Lastly, future interventions could employ advanced LLMs integrated with text classification functionalities for a more organic and contextually appropriate humor insertion, optimizing timing and effectiveness in the process. This approach would not only address the limitations of pre-defined humor placements but also contribute valuable insights for developing more engaging humor-augmented BCTs for the conversational agents for behavioral change.


\section{Conclusion}
The work underscores the potential of affiliative humor as a valuable tool for enhancing user engagement and adherence to health behavior interventions, spotlighting its applicability in healthcare interactions via the conversational agents. Implementing affiliative humor into conversational agents proves worthwhile, focusing on user engagement that directly influences intervention outcomes. 
This study represents a pioneering effort to explore and validate the effectiveness of incorporating the humorous communication style within the chatbot-delivered intervention for physical activity. 
While the content of the intervention is crucial, it becomes effective only when users find the experience engaging. Although the incorporation of humorous communication did not significantly influence users’ motivation to alter physical activity levels, aligning intervention objectives with individual intrinsic motivations is crucial for optimizing effectiveness. 
As the field is still in its nascent phase, the study has yielded promising results, making it the foremost investigation into the impact of a humorous communication style in a chatbot intervention spanning an extended duration.


\newpage




\bibliographystyle{ACM-Reference-Format}
\bibliography{sample-base}


\end{document}